\documentclass[aps,prr,twocolumn,reprint,superscriptaddress]{revtex4-2}

\usepackage{hyperref}
\usepackage{graphicx}  
\usepackage{bm}        
\usepackage{amssymb}   
\usepackage{xspace}
\usepackage{verbatim}
\usepackage{color}
\usepackage{soul}
\usepackage{subfigure}
\usepackage[export]{adjustbox}
\usepackage{dcolumn}
\usepackage{amsthm,stmaryrd,mathtools} 
\usepackage{txfonts}
\usepackage{braket}
\usepackage{amsmath}
\usepackage{xcolor}
\usepackage{array}
\usepackage{multirow}
\usepackage{tabularx}
\usepackage{booktabs}
\usepackage{lipsum}

\begin{document}

\title{Floquet-induced localization in long-range many-body systems}

\author{Rozhin Yousefjani}%
\email{RozhinYousefjani@uestc.edu.cn}
\affiliation{Institute of Fundamental and Frontier Sciences, University of Electronic Science and Technology of China, Chengdu 610051, China}

\author{Sougato Bose}
\email{s.bose@ucl.ac.uk}
\affiliation{Department of Physics and Astronomy, University College London, Gower Street, London WC1E 6BT, United Kingdom}

\author{Abolfazl Bayat}
\email{abolfazl.bayat@uestc.edu.cn}
\affiliation{Institute of Fundamental and Frontier Sciences, University of Electronic Science and Technology of China, Chengdu 610051, China}

\begin{abstract}
The fate of many-body localization in long-range interacting systems is not fully settled. 
For instance, the phase boundary between ergodic and many-body localized regimes is still under debate.  
Here, we use a Floquet dynamics which can induce many-body localization in a clean long-range interacting system through spatio-temporal disorder which are realized by regular operation of random local rotations. The phase diagram has been determined for two types of uniform and nonuniform long-range couplings.
Our Floquet mechanism shows more localizing power than conventional static disorder methods as it pushes the phase boundary in favor of the localized phase.
Moreover, our comprehensive long-time simulations provide strong support for obtained results based on static analysis.

\end{abstract}

\maketitle
\section{Introduction}
Many-Body Localization (MBL) is a profound concept in condensed matter physics for breaking the ergodicity principle~\cite{Rev1,papic1,papic2,papic3,Rev2,Laflorencie,Khemani,Huse1,Huse2}. 
To better understand the different aspects of MBL, several quantum information concepts~\cite{Schmidtgap,Negativity,Holevo,Hierarchy,MutualInformation,Diagonalentropy,
Numberentropy1,Numberentropy2} have been theoretically developed and several experiments on newly emerging quantum simulators have been performed~\cite{Iontrap1,Iontrap2,2DRDAtoms,CriticalRbatom,Rbatom,Netosuperconducting,superconducting1,superconducting2,
superconducting3,superconducting4,superconducting5,superconducting6,coldatoms1,coldatoms2,coldatoms3,coldatoms4,nitrogenvacancy1,
nitrogenvacancy2}.  
Most of the MBL literature have been dedicated to 1D short-range interactions, however, several fundamental problems remain open for systems with long-range couplings~\cite{LR1,LR2}. 
In fact, many interactions in nature are inherently long-range, and certain quantum simulators, e.g. ion-traps~\cite{Iontrap1,Iontrap2,Iontrap3} and Rydberg atoms~\cite{2DRDAtoms,CriticalRbatom,Rbatom}, are naturally governed by such interactions.
Long-range couplings can exist in various forms, e.g. tunneling or Ising-type interactions, which may affect the MBL physics very differently.
A key open problem in long-range MBL systems is how ergodic-MBL phase diagram changes in the presence of such long-range interactions~\cite{UAEG,PD1,PD2,PD3,PD4,PD5,PD6,PD7,PD8,PD9,PD10,PD12}. 
\\
 
Periodically driving many-body interacting systems, known as Floquet dynamics, is a well-known thermalizing  mechanism~\cite{Heating1,Heating2}, a phenomenon opposite to MBL. The fate of MBL systems under different Floquet dynamics have been studied in both theory~\cite{MBLuFlo1,MBLuFlo2,MBLuFlo3,MBLuFlo4,MBLuFlo5,MBLuFlo6,
MBLuFlo7,MBLuFlo8} and experiments~\cite{coldatoms4}. 
The results show that the MBL does not survive the Floquet dynamics unless the energy cannot be absorbed by the system due to either high frequency or large amplitude of the  driving field~\cite{MBLuFlo2,coldatoms4,RefB1,RefB2}. 
Alternatively, in Refs.~\cite{FloMBL1,FloMBL2,FloMBL3} the Floquet dynamics is designed to suppress the tunneling of the particles in a weakly disordered system to enhance the relative strength of disorder and interaction for generating an MBL phase.  
One may wonder whether it is possible to localize the evolution of a disorder-free Hamiltonian through the sequential action of local random rotations at regular time intervals. This can effectively induce spatio-temporal disorder  which may localize the system.
Apart from being fundamentally interesting, this has practical advantages too. 
In fact, inducing static disorder may result in leakage from the valid Hilbert space in superconducting quantum simulators or heating in ion-trap systems which make the simulation of deep MBL phase very challenging~\cite{Netosuperconducting}. Our mechanism does not suffer from this issue and thus is easier to be implemented on such quantum simulators.
\\

In this work, we introduce a Floquet mechanism that through generating spatio-temporal disorder
can localize a disorder-free Hamiltonian.
Using this localization mechanism, we fully determine the phase diagram of two different models, namely systems with uniform and non-uniform long-range couplings.
We show that our mechanism can induce MBL in certain long-range systems which cannot be localized by conventional static disordered Hamiltonians. 
Furthermore, through extensive long-time numerical simulations, we provide more support for our observation based on statistics of eigenstates. 

The structure of the paper is as follows. After presenting the considered model in section II and introducing our Floquet mechanism, the main results of the paper, namely, the phase diagram and our method for extracting the critical properties are discussed in section III. This section is followed by  the dynamical analysis of the MBL phase in section IV. Finally, in section V, we summarize our work.   

\section{Model}
We consider a spin-1/2 chain of $L$ particles interacting with long-range tunneling and Ising interaction
\begin{equation}\label{Eq. General_Hamiltonian}
H = -\sum_{i \neq j}\bigg\{\dfrac{J_x}{|i - j|^{a}}(S_i^{x}S_j^{x}+S_i^{y}S_j^{y})+\dfrac{J_z}{|i - j|^{b}}S_i^{z}S_j^{z} \bigg\}.
\end{equation}
Here $S_i^{(x,y,z)}$ is the spin-$1/2$ operator for
qubit at site $i$, $J_x{=}J_z{=}1$ are the interaction strengths,  and $a,b{>}0$ are the power-law exponents which determine the range of the tunneling and interaction, respectively. By varying these exponents one can tune the interaction geometry from a fully connected graph (for exponents being $0$) to a local nearest-neighbor 1D chain (when the exponents tend to $\infty$). 
Many types of long-range models~\cite{LR1,LR2} such as Coulomb, van der Waals, and dipole-dipole interactions, are special cases of Hamiltonian $H$ which can now be realized in ion trap~\cite{LRIT1,LRIT2} systems.  
Here, we systematically investigate two different regimes: (i) uniform couplings in which $a{=}b{\ll}\infty$;  and (ii) nonuniform couplings in which $b$ is finite (power-law Ising interaction) and $a{\rightarrow}\infty$ (nearest-neighbor tunneling).
\\

To dynamically localize this disorder-free Hamiltonian, we propose a Floquet dynamics in which the evolution over a single time period $\tau$ consists of two operations. First, the system evolves under the action of the disorder-free Hamiltonian $H$ for a short-time period $\tau$. 
Second, an instant kick operation which is a set of local random rotations along the $\hat{z}$ axis, i.e. $\mathcal{R}(\boldsymbol{\theta}){=}\prod_{i=1}^{L}e^{-i\theta_i S_i^{z}}$, rotates all the qubits without creating spin excitations, although it will induce energy excitations,  in the system. 
Since $\theta_i$ is site-dependent, these local rotations induce spatio-temporal disorder in the dynamics of the system. The random angles $\boldsymbol{\theta}=(\theta_1,\cdots,\theta_L)$ are  drawn from a uniform distribution $[-\theta/2,\theta/2]$ with $0\leq\theta/\pi\leq 1$  being the strength of the kick. Hence, the evolution operator over a single period becomes
\begin{equation}\label{Eq. floquet_dynamic}
U_F(\boldsymbol{\theta},\tau,a,b) = \mathcal{R}(\boldsymbol\theta)e^{-iH\tau}. 
\end{equation}
The random angles $\boldsymbol{\theta}$ remain fixed in different periods. 
The dynamics of the system is described  
by an effective Hamiltonian $H_F$ such that $U_F=e^{-iH_F\tau}$. 
The random nature of $\boldsymbol{\theta}$ prevents us from analytically driving a closed form for $H_F$ restricting us to numerical simulations.
Note that, our Floquet mechanism is fundamentally different from Floquet dynamical decoupling methods~\cite{FloMBL1,FloMBL2,FloMBL3} in which the suppression of the hopping amplitude increases the relative strengths of disorder and interactions that potentially drive an ergodic system toward the MBL phase.

\section{Phase diagram} 
\begin{figure}[t!]
\includegraphics[width=0.9\linewidth]{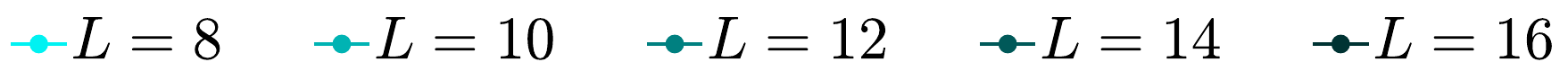}
\includegraphics[width=\linewidth]{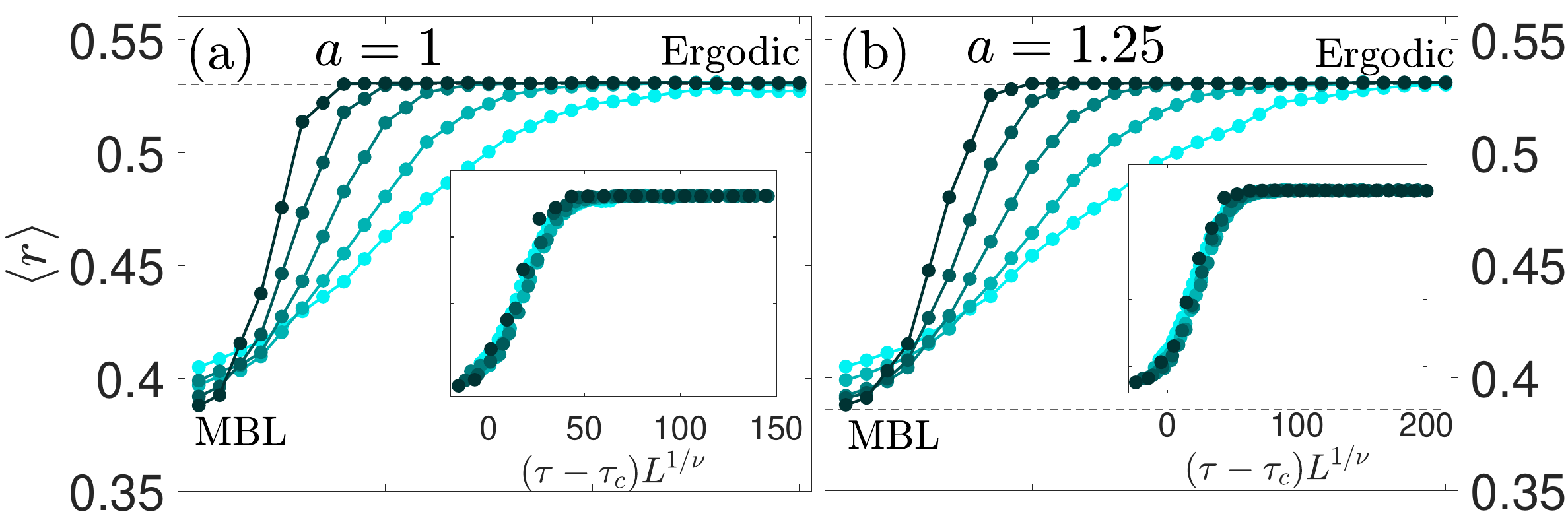}
\includegraphics[width=\linewidth]{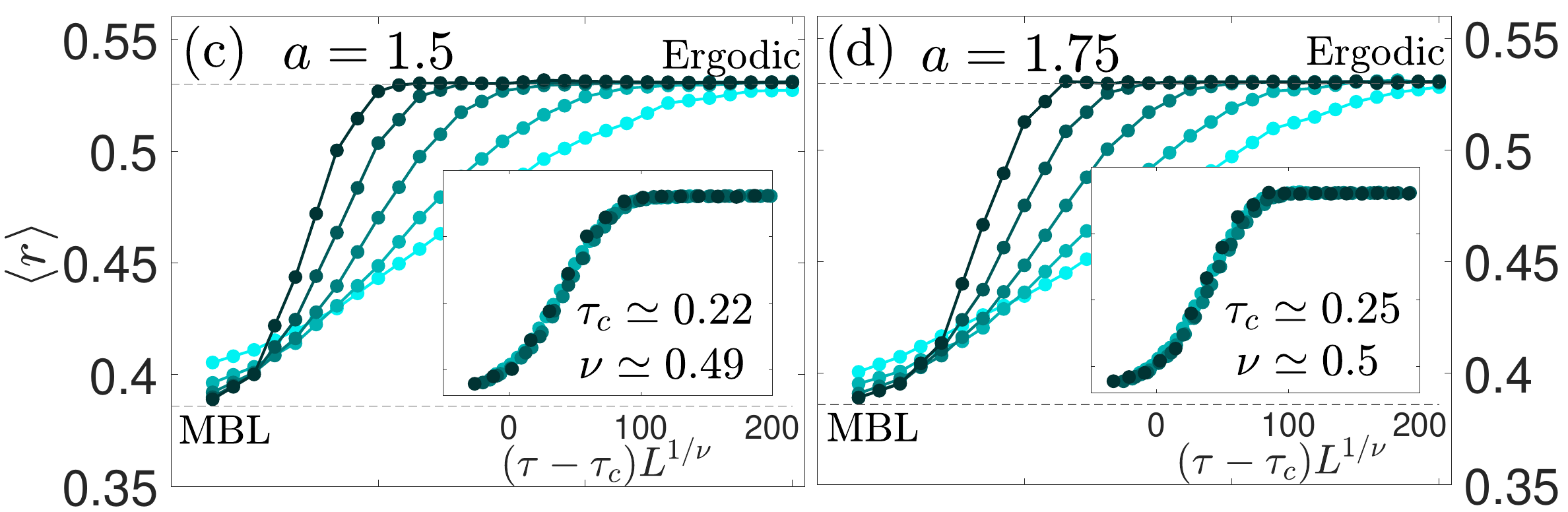}
\includegraphics[width=\linewidth]{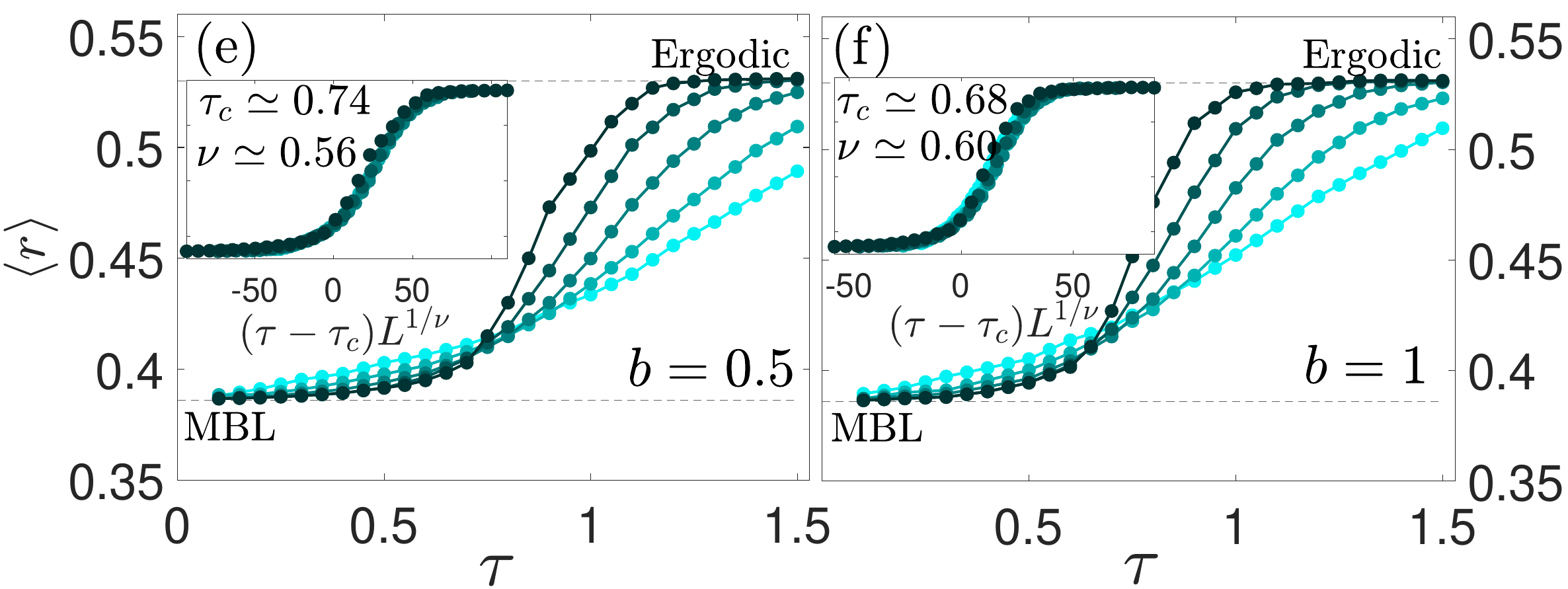}
\caption{\textbf{Upper panels:}  $\langle r\rangle$ as a function of $\tau$ for uniform couplings: (a) $a{=}b{=}1$; (b) $a{=}b{=}1.25$; (c) $a{=}b{=}1.5$; and (d) $a{=}b{=}1.75$. 
In all panels, the results are obtained for fixed $\theta{=}\pi$ in systems with various sizes and the dashed lines correspond to $\langle r\rangle{\simeq} 0.529$ and $\langle r\rangle{\simeq} 0.386$ for ergodic and MBL phases, respectively.
The insets are the best data collapse obtained through the finite-size scaling analysis for extracting $\tau_c$ and $\nu$.
For any choice of $a{\geq} 1.5$, our finite-size scaling analysis unambiguously determines the transition point $\tau_c$ and critical exponent $\nu$.
Since the quality of data collapse drops significantly for $a{<}1.5$, despite the fact that $\langle r \rangle$ is close to $0.4$ for small $\tau$, which is possibly due to finite-size effects, no phase transition is detected for the considered system sizes. 
This lack of phase transition indicates that there is no localized phase for $a{<}1.5$. 
\textbf{Lower panels:} 
$\langle r\rangle$ as a function of $\tau$ for nonuniform couplings (i.e. $b{\ll}a{\rightarrow}\infty$): (e) $b{=}0.5$; and (f) $b{=}1$.
In both panels, we set $\theta{=}\pi$.
The insets are the best data collapse obtained through the finite-size scaling analysis for extracting the attached transition point $\tau_c$'s and critical exponent $\nu$'s.       
}\label{fig:LSs}
\end{figure} 
\begin{figure}[t!]
\includegraphics[width=\linewidth]{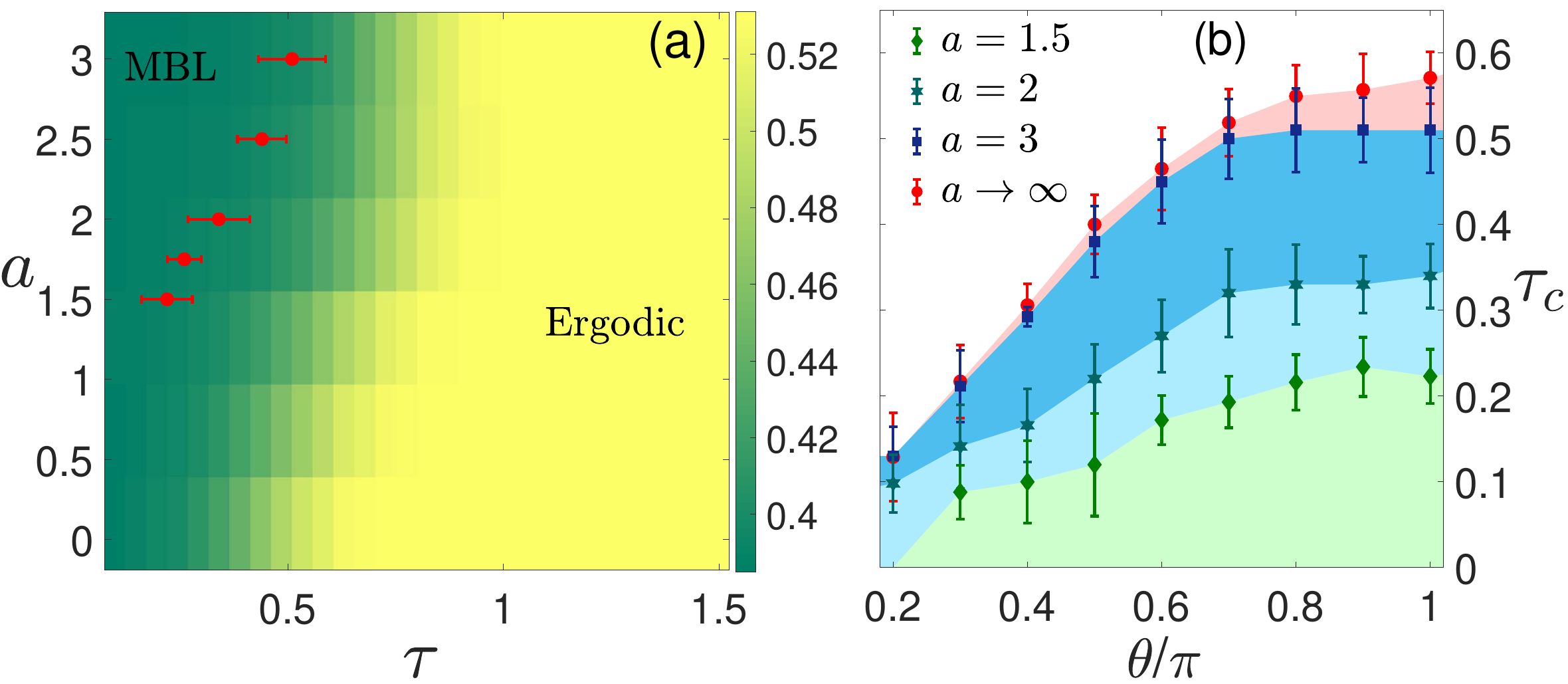}
\includegraphics[width=\linewidth]{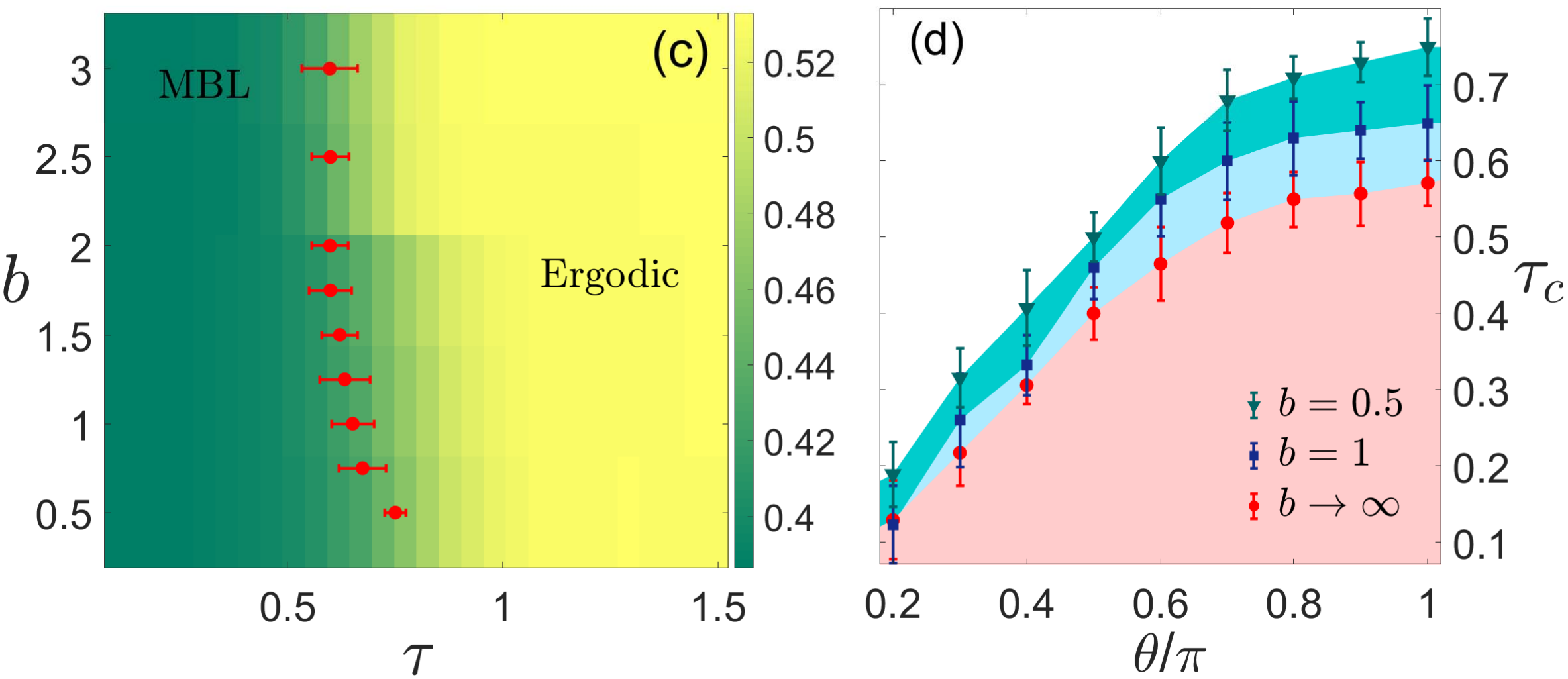}
\caption{\textbf{Left panels:}
$\langle r\rangle$ for a system of length $L=16$ and $\theta=\pi$ as a function of $\tau$ and power-law exponents for (a) uniform couplings ($a=b$); and (c) nonuniform couplings ($b{\ll}a{\rightarrow\infty}$). 
In both panels, the red markers show the MBL-ergodic phase boundary determined by the finite-size scaling analysis.
\textbf{Right panels:} the critical $\tau_c$ as a function of $\theta$ for various choices of exponents in (b) uniform couplings ($a=b$); and (d) nonuniform couplings ($b{\ll}a{\rightarrow\infty}$).
In both panels, the area below each curve represents the MBL phase.
}\label{fig:PDs}
\end{figure}

Since the dynamics of the system is described by the effective Hamiltonian $H_F$, one can investigate the properties of this Hamiltonian to determine the phase diagram of the system as a function of parameters $(\theta,\tau,a,b)$. 
We first investigate the statistical properties of energy levels of $H_F$, namely $\{ E_k\}$, or equivalently the quasi-energy levels of $U_F$ given by $\{ e^{-iE_k\tau}\}$. Numerically, we compute the quasi-energies using exact diagonalization of $U_F$ in the subspace of $S^z_{tot}{=}0$, with $S^z_{tot}{=}\sum_j S_j^z$. All the results are averaged  over $1000$ random samples of $\boldsymbol{\theta}$ to guarantee proper convergence. 
Since the Hamiltonian in Eq.(\ref{Eq. General_Hamiltonian}) and its corresponding $H_F$ are dramatically less sparse than the nearest-neighbor interactions, the numerical simulation of the system is very challenging~\cite{MPS} and thus is restricted to $L{=}16$. 
By computing the consecutive energy gaps $\delta_{k}{=}E_{k+1}{-}E_{k}$, one can characterize the level statistics by their ratio $r_{k}{=}\min(\delta_{k+1},\delta_{k}){/}\max(\delta_{k+1},\delta_{k})$.
The averaged value of this ratio, $\langle r\rangle$, serves as a well-established tool in numerical studies of finite MBL systems~\cite{LSR1}. 
While the MBL phase is determined by Poisson level statistics with $\langle r\rangle{\simeq} 0.386$, the ergodic phase is known to follow the circular orthogonal ensemble level statistics with $\langle r\rangle{\simeq} 0.529$~\cite{LSR2,Laflorencie}.
\\ 

Clearly, we have four control parameters in the system, namely $(\theta,\tau,a,b)$. 
For each choice of these parameters, one can compute $\langle r\rangle$ to reveal the phase of the system. 
As an example, by fixing $\theta{=}\pi$ and considering various system sizes, in Figs.~\ref{fig:LSs}(a)-(d), we plot $\langle r\rangle$ as a function of $\tau$ for several choices of uniform couplings $a{=}b{\in}\{1,1.25,1.5,1.75\}$. 
Two main features can be observed. 
First, as $\tau$ increases the system becomes ergodic and this transition gets sharper by increasing the system size. 
This can be understood as by increasing $\tau$ 
the evolution of the disorder-free Hamiltonian $H$ gets enough time to thermalize the system. 
In other words, the effective Hamiltonian $H_F$ is dominantly determined by the Hamiltonian $H$ rather than the disordered kick operator $\mathcal{R}(\boldsymbol\theta)$. 
This is fundamentally different from those Floquet systems in which energy absorption is suppressed due to the high frequency of the ergodic evolution which makes static disorder dominant~\cite{MBLuFlo1,MBLuFlo2,MBLuFlo3,MBLuFlo4,MBLuFlo5,MBLuFlo6,
MBLuFlo7,MBLuFlo8}.  
Second, for any choice of $a{\geq}1.5$ (Fig.~\ref{fig:LSs} (c) and (d)) the curves of different sizes clearly intersect in tiny domains of $\tau$, which indicates the emergence of the scale invariance in the vicinity of the transition point $\tau_c$.
Note that the intersection domains for the curves in Fig.~\ref{fig:LSs} (a) and (b), i.e. for the case $a{\leq}1.5$, increase by adding system sizes which signaling disorder-dependent transition points~\cite{PD7,PD9}. 
In the context of MBL, the scale invariance
implies the emergence of a diverging length scale $\xi$ in the system and thus scaling the interested observable as 
$\mathcal{F}(L/\xi)$.
Here, $\mathcal{F}$ is an arbitrary function and $L$ denotes the system size.
Considering ergodic-MBL transition as a continuous
second-order transformation results in the diverging length scale  $\xi{\propto}|\tau-\tau_c|^{-\nu}$ with $\nu$ as  the critical exponent.
Precise determination of transition point $\tau_c$ and critical exponent $\nu$ demand finite-size scaling analysis.
To do this, we plot $\langle r\rangle$ as a function of $(\tau{-}\tau_c)L^{1/\nu}$. 
By proper choice of $\tau_c$ and $\nu$, one can collapse the curves for different system sizes.
To achieve the best
data collapse we use an elaborate optimization scheme
and minimize a proper quality function $Q$~\cite{pyfssa1,pyfssa2,Qfunction},
which is defined and discussed in the Appendix. In our case, a perfect data collapse results in  $Q{=}1$ and any deviation from such a perfect situation makes $Q$ larger. 
Results of finite-size scaling analysis are shown in the inset of Figs.~\ref{fig:LSs}(a)-(d) for uniform couplings $a{=}b$.
While, for any choice of $a{\geq} 1.5$, the finite-size scaling unambiguously determines the transition point $\tau_c$ and critical exponent $\nu$ with $Q\sim 4$, the quality of data collapse drops significantly for $a{<}1.5$.
The smallest $Q$'s for the corresponding data collapses are obtained about $200$, for any choice of $\tau_c$ and $\nu$ and considered system sizes.
Therefore, for $a{<}1.5$ and considered system sizes, despite the fact that $\langle r \rangle$ is close to $0.4$ for small $\tau$, which is possibly due to finite-size effects, one cannot confidently find scaling behavior which is expected near the transition point and in the thermodynamic limit, there will be no localized phase in this regime. 
This is an interesting observation as the power-law couplings $a$ and $b$ play two opposite roles. Decreasing the coupling $a$ allows spin tunneling between the distant qubits which enhances ergodicity. 
On the other hand, decreasing the coupling $b$ (i.e. making the Ising interaction more long-range) creates an effective site-dependent energy shift whose value depends on the spin configuration of the whole system. This energy shift acts like  an effective random magnetic field which enhances localization. 
The absence of MBL in uniform couplings (i.e. $a{=}b$) for $a{<}a_c{\simeq}1.5$ shows that the thermalizing long-range tunneling  overcomes the localizing long-range Ising. This has also been observed in ordinary disordered long-range Hamiltonians. However, while in such systems $a_c$ is found to be $a_c{\simeq} 2$~\cite{PD3,Safavi,UAEG}, our Floquet system shows more localization power with $a_c\simeq 1.5$. 
In other words, the spatio-temporal disorder $\mathcal{R}(\boldsymbol\theta)$ has more localization power than the spatial one and gives $\tau_c{=}0.22$ and $\tau_c{=}0.25$ for $a{=}1.5$ and $a{=}1.75$, respectively. 
To investigate the phase diagram of the uniform case (i.e. $a{=}b$) with more details, we keep the strength of the random kick to a strong value of $\theta{=}\pi$ and plot $\langle r \rangle$ as a function of $a$ and $\tau$ in Fig.~\ref{fig:PDs}(a) for a system of size $L{=}16$. The boundary between the ergodic and the MBL phases, denoted by red markers, is determined by finite-size scaling analysis of $\langle r \rangle$, as discussed before.  
In addition, to clarify the role of random kick strength, in Fig.~\ref{fig:PDs}(b) we plot the critical time $\tau_c$  as a function of $\theta$ for various $a$'s. 
The area below each curve represents the MBL phase. 
Clearly, by reducing $a$ the MBL area shrinks, showing the tendency toward thermalization. 
\\

All the above analysis can be repeated for systems with nonuniform couplings ($b{\ll} a{\rightarrow} \infty$).
Again as an example, by fixing $\theta{=}\pi$ and considering various system sizes, in Figs.~\ref{fig:LSs}(e)-(f), we plot $\langle r\rangle$ as a function of $\tau$ for two choices of nonuniform couplings $b{\in}\{0.5,1\}$.
For considered system sizes, one can see the clear intersection points for all the curves in  Figs.~\ref{fig:LSs}(e)-(f) which determine the onset of transition. The finite-size scaling analysis which collapses all the curves on a universal one as a function of $(\tau{-}\tau_c)L^{1/\nu}$ and leads to precise $\tau_c$ are presented in the insets of Figs.~\ref{fig:LSs} (e) and (f). 
The achieved $Q$'s for these data collapses are about $4$. 
As long-range Ising interaction induces an effective static disorder in the chain, the free evolution of the clean Hamiltonian $H$ takes a longer time to thermalize the system.
Therefore the transition points are highly skewed to the larger values of $\tau$.   
We obtain $\tau_c{=}0.74$ and $\tau_c{=}0.68$ for systems with nonuniform couplings $b{=}0.5$ and $b{=}1$, respectively.           
To determine the whole phase diagram of the system for  nonuniform couplings, in Fig.~\ref{fig:PDs} (c), we plot $\langle r \rangle$ as a function of $b$ and $\tau$ in a system of size $L{=}16$, when the strength of the disordered kick is fixed to $\theta{=}\pi$.
The phase boundary between the MBL and the ergodic phases in Fig.~\ref{fig:PDs}(b) are denoted by red markers. 
By decreasing $b$ (i.e. making the Ising interaction more long-range) the localization power enhances and the system can localize for longer $\tau$.
As discussed above, decreasing $b$ induces effective disorder and thus $\tau_c$ is increased.    
To see the effect of disordered kick strength on the phase diagram, in Fig.\ref{fig:PDs} (d) we plot $\tau_c$ as a function of $\theta$ for various  $b$'s. The area below the curves represents the MBL phase. As expected, by decreasing $b$ the MBL region increases which further confirms the enhancement of the localization power. 

\section{Dynamical analysis of the MBL phase} 
\begin{figure}[t!]
\includegraphics[width=\linewidth]{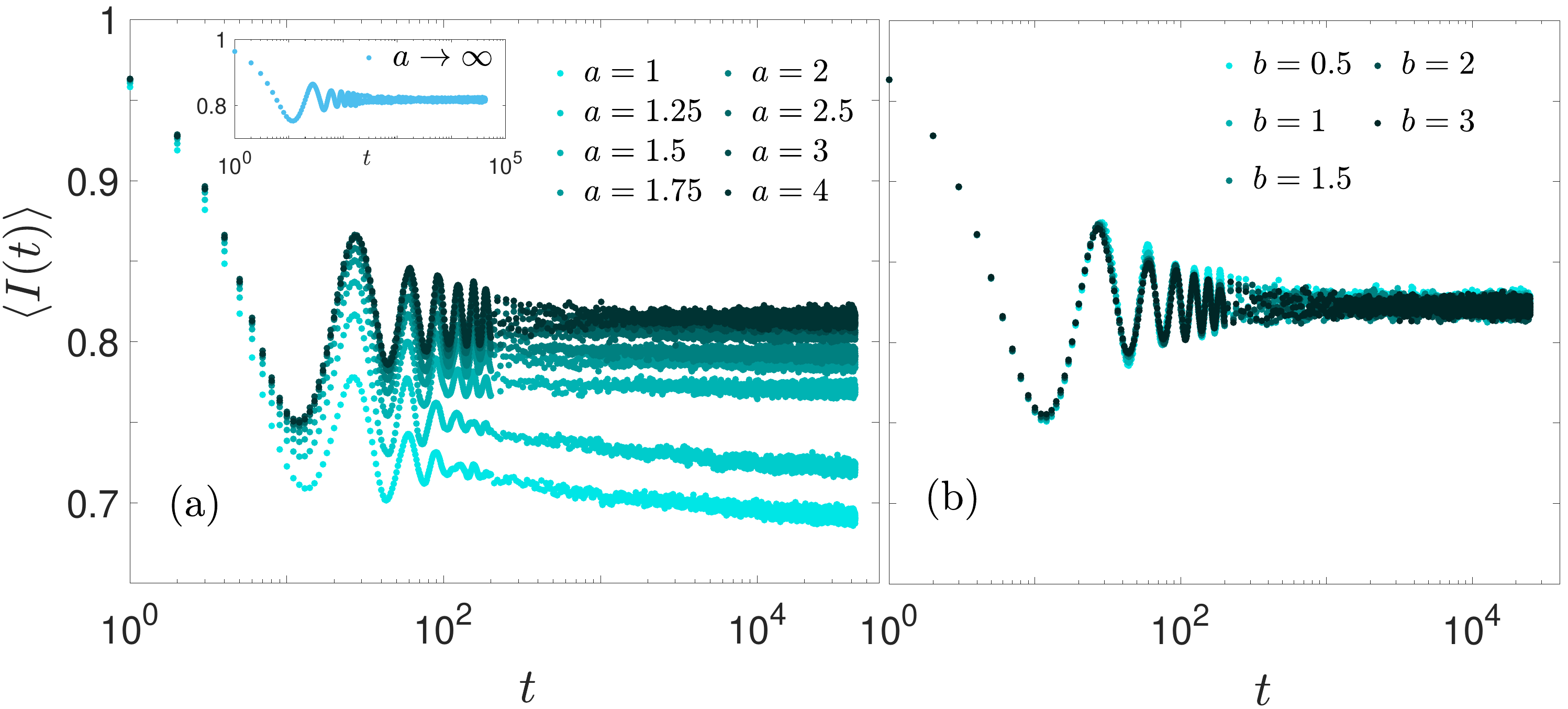}
\caption{ Imbalance $\langle I(t)\rangle$ versus $t$ in the MBL phase for $\theta{=}\pi$, $\tau{=}0.1$ and $L{=}16$. Panel (a) and its inset are for the uniform couplings ($a{=}b$).  Panel (b) is for the nonuniform couplings ($b{\ll}a{\rightarrow}\infty$). }\label{fig:MI}
\end{figure} 
To better understand the MBL phase in long-range interacting systems, it is highly insightful to investigate the dynamical properties.
Here, we focus on measuring imbalance that can quantify the ability of the system to conserve the initial information.
In fact, we pursue two main objectives: (i) illustrating the localization dynamics of the system; and (ii) providing further affirmation for the phase boundary determined by $\langle r\rangle$, in particular the absence of MBL for $a{=}b{<}1.5$.
\\

We initialize a system of size $L{=}16$ in N\'eel state $\vert\psi(0)\rangle{=}\vert{\uparrow\downarrow\ldots\uparrow\downarrow}\rangle$. 
The evolution of the system  after $t$ times kicking is given by 
$\vert\psi(t)\rangle{=} (U_F)^t \vert\psi(0)\rangle$. 
The imbalance is defined as $I(t){=}2/L\sum_{i}(-1)^{i+1}\langle S^{z}_{i}\rangle$, where $\langle S^{z}_{i}\rangle{=}\langle \psi(t) \vert S^{z}_{i}\vert\psi(t) \rangle$ and the normalization in the definition guarantees that  $I(0){=}1$.
In the following, we set $\theta{=}\pi$ and $\tau{=}0.1$ to be sure that the system evolves in MBL phase.
For achieving good statistics and converging results, we generate 1000 random samples and denote the random-averaged imbalance as $\langle I(t) \rangle$.
While in the ergodic phase, the imbalance has to relax to zero, showing no memory about the initial state, in the MBL phase it reaches a finite value, resembling the presence of memory~\cite{coldatoms1,coldatoms4,Hierarchy}. Fig.~\ref{fig:MI} (a) illustrates random-averaged imbalance $\langle I(t) \rangle$ versus $t$ for vrious values of uniform couplings ($a{=}b$). 
After a transport time, the imbalance relaxes to a plateau for  $a{\geq}1.5$, signaling that the system is strongly localized and all particles will stay close to their original positions during time evolution. 
This is in full agreement with the level statistics analysis presented in the previous section. 
For $a{<}1.5$ the imbalance gradually relaxes to zero and, hence, the system will thermalize in a long-time.
This confirms that for the choice of $\tau{=}0.1$, $\theta{=}\pi$ the critical power-law coupling is $a_c{\simeq} 1.5$, again in agreement with level statistics analysis.
For the sake of completeness, in the inset of  Fig.~\ref{fig:MI} (a) we plot the random-averaged imbalance for a system with short-range tunneling and Ising interaction, i.e. $a{=}b{\rightarrow}\infty$. 
For nonuniform couplings, in  Fig.~\ref{fig:MI}(b), we plot the imbalance as a function of $t$ for various $b$'s. 
Interestingly, in the localized phase the dynamics of imbalance and its saturation hardly changes by $b$. 
Clearly, all the curves after some transport time relax to a non-zero constant showing that system can preserve the initial information during the long-time simulation.
       
\section{Conclusion}       
We have proposed a Floquet mechanism, which enables the creation of the MBL phase in a disorder-free long-range interacting system.
In the limit of short-time evolution $\tau$, the mechanism reproduces the results for conventional disordered systems. 
By utilizing this Floquet mechanism, two main results have been achieved. Firstly, we have determined the phase diagram of the system for two different types of couplings, namely uniform ($a{=}b$) and nonuniform ($b{\ll} a{\rightarrow} \infty$), using level statistics. 
Our mechanism shows a strong localizing power such that it prevents thermalization in those long-range systems which cannot be localized merely by disorder. 
Secondly, we have studied the dynamics of imbalance to provide further support for the level-statistics analysis in both types of couplings.

\section{Acknowledgments} 
AB acknowledges support from the National Key R\&D Program of China (Grant No. 2018YFA0306703), the National Science Foundation of China (Grants No. 12050410253, No. 92065115, and No. 12274059), and the Ministry of Science and Technology of China (Grant No. QNJ2021167001L).
SB acknowledges the EPSRC grant for nonergodic quantum manipulation (Grant No.~EP/R029075/1). RY acknowledges the National Science Foundation
of China for the International Young Scientists Fund
(Grant No. 12250410242).

\appendix* 
\setcounter{equation}{0}
\setcounter{figure}{0}
\setcounter{table}{0}
\renewcommand{\theequation}{A\arabic{equation}}
\renewcommand{\thefigure}{A\arabic{figure}}
\renewcommand{\thetable}{A\arabic{table}}
\section{} 
To precisely determine the critical values, we perform finite-size scaling analysis using Python package pyfssa and evaluate the quality of the data collapse as follows. Assuming that $i$ indexes the system size $L_i$ and $j$ indexes the time period $\tau_{j}$ with $\tau_{1}<\tau_2<\cdots<\tau_k$. 
For scaled observations $\{y_{ij}\}$ (e. g. the random-averaged level statistics ratio $\langle r\rangle$) and its standard errors  $\{dy_{ij}\}$ at $x_{ij}=L_{i}^{1/\nu}(\tau_{j} -\tau_c)$, and also $\{\hat{y}_{ij}\}$ and $\{d\hat{y}_{ij}\}$ as the estimated values of the master curve  and its standard error again at $x_{ij}=L_{i}^{1/\nu}(\tau_{j} -\tau_c)$, Houdayer, and Hartmann [86] redefined the quality function 
\begin{equation}
Q=\dfrac{1}{\mathcal{N}}\sum_{ij}\dfrac{(y_{ij} - \hat{y}_{ij})^2}{dy_{ij}^2 - d\hat{y}_{ij}^2}.
\end{equation}
The sum in the quality function $Q$ only involves terms for which the estimated value $\hat{y}_{ij}$ of the master curve at $x_{ij}$ is defined. The number of such terms is $\mathcal{N}$. 
For an optimal fit, the individual deviations $(y_{ij} - \hat{y}_{ij})^2$   is of the order of the individual error $dy_{ij}^2 - d\hat{y}_{ij}^2$, so the quality $Q$ is close to $1$ and much larger for non-optimal fits.

%

\end{document}